\documentclass[floatfix,twocolumn,nofootinbib]{revtex4}

\usepackage[utf8]{inputenc}
\usepackage{amsfonts,amsmath,amssymb}
\usepackage{graphicx}
\usepackage{xspace}
\usepackage{mathtools}
\usepackage[caption=false]{subfig}
\usepackage{hyperref}
\usepackage{bm}
\usepackage[export]{adjustbox}

\let\oldcite\cite
\renewcommand{\cite}{\unskip~\oldcite}
\newcommand{\reftable}[1]{Table~\ref{#1}}
\newcommand{\reffig}[1]{Fig.~\ref{#1}}
\renewcommand{\refeq}[1]{Eq.~(\ref{#1})}
\newcommand{\refsec}[1]{Sec.~\ref{#1}}
\newcommand{\refcite}[1]{Ref.\cite{#1}}
\newcommand{\refapp}[1]{Appendix~\ref{#1}}
\newcommand{\see}[1]{(see \eg\refcite{#1})}

\newcommand{\gev}{\ensuremath{\,\text{GeV}}}
\newcommand{\tev}{\ensuremath{\,\text{TeV}}}
\newcommand{\invfb}{\ensuremath{/\text{fb}}}

\newcommand{\roots}[1]{\ensuremath{\sqrt{s}=#1\tev}}
\newcommand{\s}[1]{\ensuremath{#1\tev}}

\let\olddigamma\digamma
\renewcommand{\digamma}{\ensuremath{\olddigamma}\xspace}
\newcommand{\md}{\ensuremath{m_\digamma}}
\newcommand{\gd}{\ensuremath{\Gamma_\digamma}}

\newcommand{\eg}{e.g.,\xspace}
\newcommand{\ie}{i.e.,\xspace}

\DeclarePairedDelimiter\parentheses{\lparen}{\rparen}

\newcommand{\probsymbol}{\ensuremath{p}}
\newcommand{\prob}[1]{\ensuremath{\probsymbol\parentheses*{#1}}}
\newcommand{\cond}[2]{\ensuremath{\prob{#1 \bm\mid #2}}}

\newcommand{\data}{\text{ATLAS data}}
\newcommand{\SM}{\text{SM}}
\newcommand{\SMdigamma}{\text{SM + \digamma}}

\newcommand{\like}{\mathcal{L}}

\begin{document}

\title{Bayes-factor of the ATLAS diphoton excess}

\author{Andrew Fowlie}
\affiliation{ARC Centre of Excellence for Particle Physics at the Tera-scale, School of Physics and Astronomy, Monash University, Melbourne, Victoria 3800 Australia}

\date{\today}

\begin{abstract}
We present a calculation of Bayes-factors for the digamma resonance (\digamma) versus the SM in light of ATLAS \s{8} $20.3\invfb$, \s{13} $3.2\invfb$ and \s{13} $15.4\invfb$ data, sidestepping any difficulties in interpreting significances in frequentist statistics. We matched, wherever possible, parameterisations in the ATLAS analysis. We calculated that the plausibility of the \digamma versus the Standard Model increased by about eight in light of the \s{8} $20.3\invfb$ and \s{13} $3.2\invfb$ ATLAS data, somewhat justifying interest in \digamma models. All told, however, in light of $15.4\invfb$ data, the \digamma was disfavoured by about $0.7$.
\end{abstract}

\maketitle

\section{Introduction}

The statistical anomalies at about $750\gev$ in ATLAS\cite{atlas,Aaboud:2016tru} and CMS\cite{CMS:2015dxe,Khachatryan:2016hje} searches for a diphoton resonance (denoted in this text as \digamma) at \roots{13} with about $3\invfb$ caused considerable
activity \see{Ellis:2015oso,Franceschini:2015kwy,Strumia:2016wys}. The experiments reported local significances, which incorporate a look-elsewhere effect (LEE, see \eg \refcite{Lyons:2013yja,2008arXiv0811.1663L}) in the production cross section of the \digamma, of $3.9\sigma$ and $3.4\sigma$,
respectively, and global significances, which incorporate a LEE in the production cross section, mass and width of the \digamma, of $2.1\sigma$ and $1.6\sigma$, respectively. There was concern, however, that an overall LEE, accounting for the numerous hypothesis tests of the SM at the LHC, cannot be incorporated, and that the plausibility of the \digamma was difficult to gauge. 

Whilst ultimately the \digamma was disfavoured by searches with about $15\invfb$\cite{CMS:2016crm,ATLAS:2016eeo}, we directly calculate the relative plausibility of the SM versus the SM plus \digamma in light of ATLAS data available during the excitement, matching, wherever possible, 
parameter ranges and parameterisations in the frequentist analyses. The relative plausibility sidesteps technicalities about the LEE and the frequentist formalism required to interpret significances. We calculate the
Bayes-factor \see{Gregory} in light of ATLAS data,
\begin{equation}
\frac{\cond{\data}{\SMdigamma}}{\cond{\data}{\SM}}
=
\frac{
\frac{\cond{\SMdigamma}{\data}}{\cond{\SM}{\data}}
}{ 
\frac{\prob{\SMdigamma}}{\prob{\SM}}
}.
\end{equation}
Our main result is that we find that, at its peak, the Bayes-factor was about $7.7$ in favour of the \digamma. 
In other words, in light of the ATLAS \s{13} $3.2\invfb$ and \s{8} $20.3\invfb$ diphoton searches, the relative plausibility of the \digamma versus the SM alone 
increased by about eight. This was ``substantial'' on the Jeffreys' scale\cite{Jeffreys:1939xee}, lying between ``not worth more than a bare mention'' and ``strong evidence.'' For completeness, we calculated that this preference was reversed by the ATLAS \s{13} $15.4\invfb$ search\cite{ATLAS:2016eeo}, resulting in a Bayes-factor of about $0.7$. Nevertheless, the interest in \digamma models in the interim was, to some degree, supported by Bayesian and frequentist analyses. Unfortunately, CMS performed searches in numerous event categories, resulting in a proliferation of background nuisance parameters and making replication difficult without cutting corners or considerable
computing power.

\section{Calculation}

The background shape was characterised by a monotonically decreasing function with two free parameters,
\begin{equation}\label{eq:bg_dist}
p_b(m_{\gamma\gamma}) \propto \left[1 -  \left(\frac{m_{\gamma\gamma}}{\sqrt{s}}\right)^{\tfrac13}\right]^b \left(\frac{m_{\gamma\gamma}}{\sqrt{s}}\right)^{a}.
\end{equation}
The \roots{8} and \roots{13} backgrounds were described by separate choices of $a$, $b$ and normalisation, $n_b$. 

ATLAS modelled the experimental resolution of the signal shape with a double-sided crystal ball (DSCB) function\cite{atlas,Aaboud:2016tru}. In their combined analysis\cite{atlas}, ATLAS accounted for a substantial width by promoting DSCB parameters to functions of the mass and width of the \digamma. Because the details of this treatment were not published, we picked a simpler ansatz for the signal shape and experimental resolution. The \digamma signal was described by a Breit-Wigner or a Gaussian with a width equal to the ATLAS diphoton resolution if the width, $\gd$, was narrower than the ATLAS diphoton resolution, $\sigma$,
\begin{equation}\label{eq:signal_dist}
p_s(m_{\gamma\gamma})  \propto
\begin{dcases} 
    \frac{1}{(m_{\gamma\gamma}^2 - \md^2)^2 + \gd^2 \md^2} & \gd > \sigma \\
    e^{-\frac{(m_{\gamma\gamma} - \md)^2}{2\sigma}} & \gd \le \sigma
\end{dcases},
\end{equation}
and normalisation factors, $n_s$, at \s{8} and \s{13}. The ATLAS diphoton resolution was modelled by a linear function of \digamma mass,
\begin{equation}
\sigma \approx 6 \cdot 10^{-3} \md + 0.8 \gev,
\end{equation}
motivated by information in \refcite{atlas} that it changes from $2\gev$ to $13\gev$ between masses of $200\gev$ and $2\tev$. We described the normalisation factors by an expected number of signal events in the \s{13} $3.2\invfb$ search and scaling factors, reflecting the decreased cross section at \roots{8} and different integrated luminosities. Thus the SM ansatz and \digamma ansatz were described by six and four parameters, respectively.

We scraped the ATLAS bin counts, $\{n\}$, and bin edges from \refcite{Aaboud:2016tru,ATLAS:2016eeo}. To convert the distributions into expected numbers of events per bin, $\{\lambda\}$, we used analytic integration. Our likelihood function was simply a product of Poisson distributions,
\begin{equation}\label{eq:like}
\like = \prod_i \frac{\lambda_i^{n_i} e^{-\lambda_i}}{n_i!},
\end{equation}
for the expected and observed number of events in each bin in the \s{8} and \s{13} searches. In total, our calculations required about 6 million calls of our likelihood function.

%
%

\newcommand{\range}[2]{\ensuremath{\left[#1, #2\right]}}
\begin{table}
\begin{ruledtabular}
\begin{tabular}{@{\extracolsep{2pt}}llll@{}}  
\multicolumn{2}{c}{SM} & \multicolumn{2}{c}{\digamma}\\
\cline{1-2} \cline{3-4}
Parameter & Prior & Parameter & Prior\\
\cline{1-2} \cline{3-4}
$a_{13}$, $a_8$ & Flat, \range{-25}{25} & $\md / {1\tev} $ & Log, \range{0.2}{2}\\
$b_{13}$, $b_8$ & Flat, \range{-25}{25} & $\alpha$ & Log, \range{5\cdot10^{-6}}{0.1}\\
$n_{b13}$ & Log, \range{5\cdot10^3}{10^4} & $\sigma_{13/8}$ & Flat, \range{2.5}{5}\\
$n_{b8}$ & Log, \range{2\cdot10^4}{3\cdot10^4} & $n_{s13}$ & Log, \range{5}{200}\\
\end{tabular}
\end{ruledtabular}
\caption{\label{tab:priors}Priors for the SM ansatz and \digamma resonance. Subscript numbers refer to \s{8} and \s{13} $3.2\invfb$, \eg $n_{b13}$ refers to the number of expected background events at \s{13} with $3.2\invfb$.
}
\end{table}

The priors for the parameters were the final ingredients. We picked logarithmic priors for $\md$ and $\alpha\equiv\gd / \md$ that matched the ranges in the ATLAS search: a mass between $200\gev$ and $2000\gev$, and $\alpha$ between $5\cdot10^{-6}$ and $0.1$. The latter range spans the narrow-width approximation (NWA) to substantial widths. We picked a logarithmic prior for the expected number of \digamma events in the \s{13} $3.2\invfb$ search, $n_{s13}$, between $5$ and $200$ events, which reflected the \digamma models anticipated in the experimental search. The number of expected events in the \s{8} search was modelled by a scaling factor between the \s{13} and \s{8} production cross sections, $\sigma_{13/8}$. We picked a linear prior, \ie $p(\sigma_{13/8}) = \text{const.}$, between $2.5$ (corresponding to a light quark initial state) and $5$ (corresponding to gluon fusion). We, of course, included a factor reflecting the decreased integrated luminosity. Since the models were composite (that is, we considered the SM and SM plus \digamma), we anticipated limited sensitivity to the priors for the SM background ansatz. We list all priors in \reftable{tab:priors} and discuss them further in \refsec{sec:priors}.

\begin{figure}
\centering

\subfloat[Combined best-fit and spectrum at \roots{8}.]{
\includegraphics[width=0.8\linewidth]{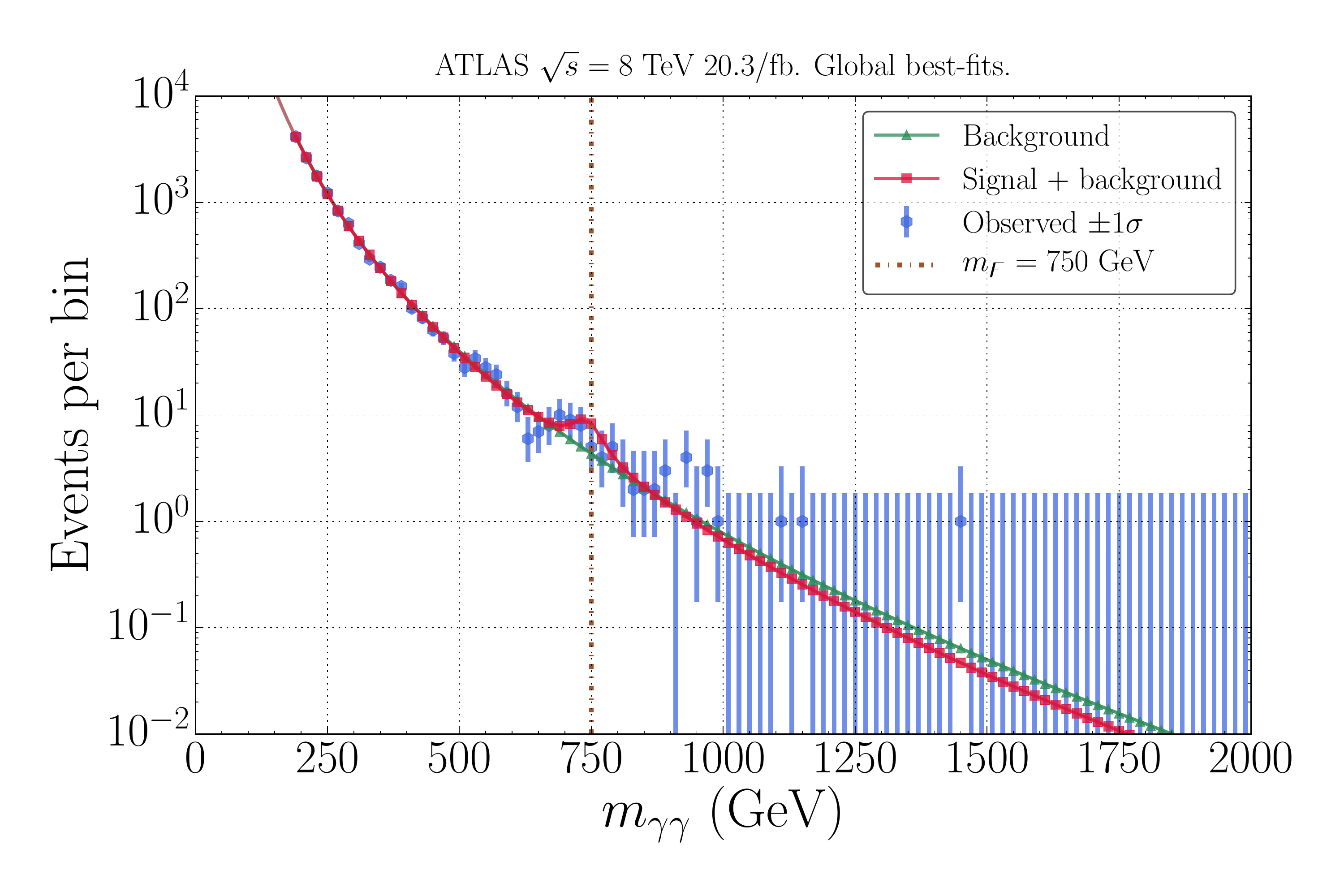}
\label{fig:digamma_8}
}

\subfloat[Combined best-fit and spectrum at \roots{13} $3.2\invfb$.]{
\includegraphics[width=0.8\linewidth]{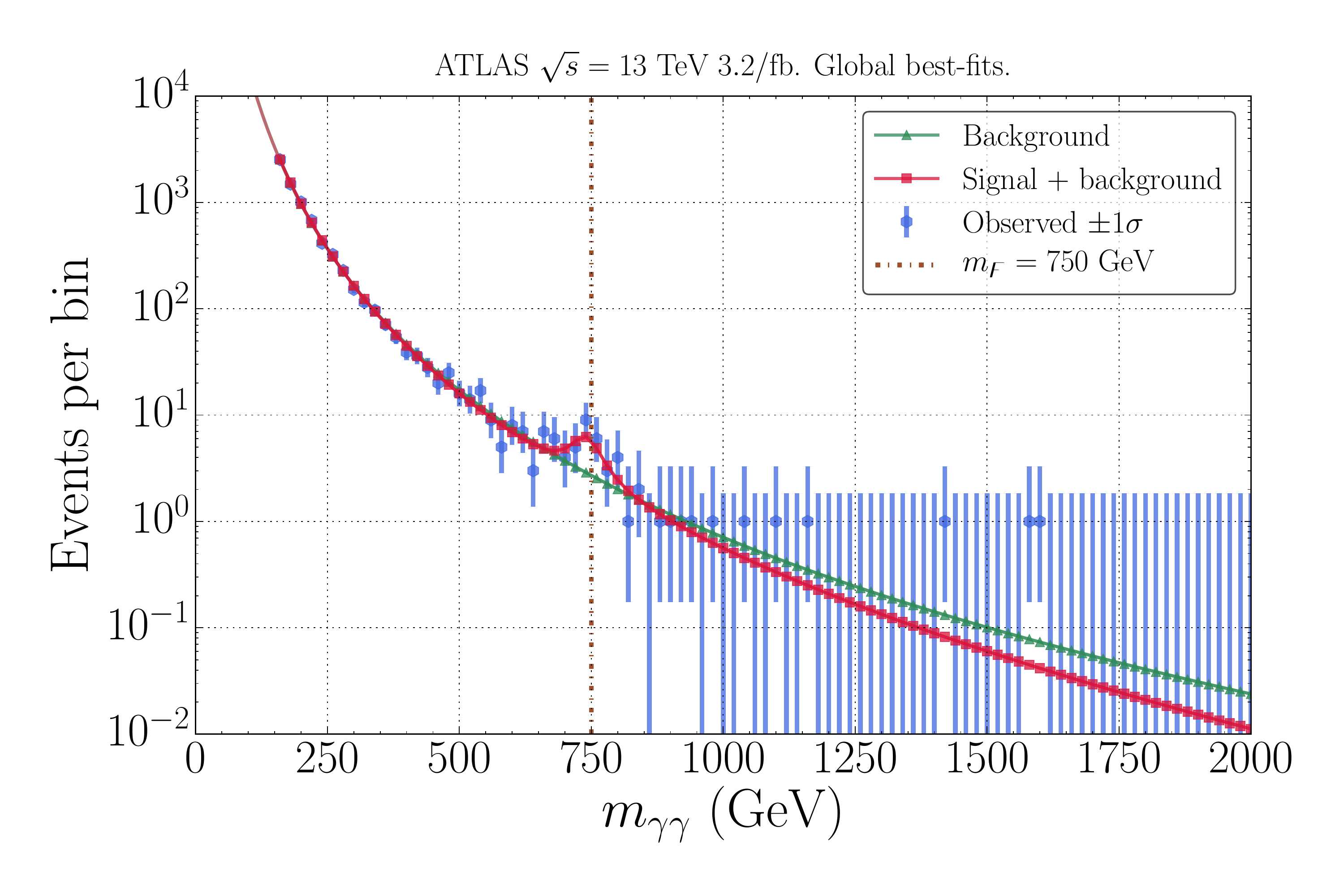}
\label{fig:digamma_13}
}%
\caption{\label{fig:best_fits} Diphoton spectrum at \protect\subref{fig:digamma_8} \roots{8} and \protect\subref{fig:digamma_13} \roots{13} $3.2\invfb$, along with best-fit background and signals.}
\end{figure}

\begin{figure*}
\centering
\subfloat[ATLAS \roots{8} $20.3\invfb$ data.]{
\setcounter{subfigure}{1}
\includegraphics[width=0.325\linewidth]{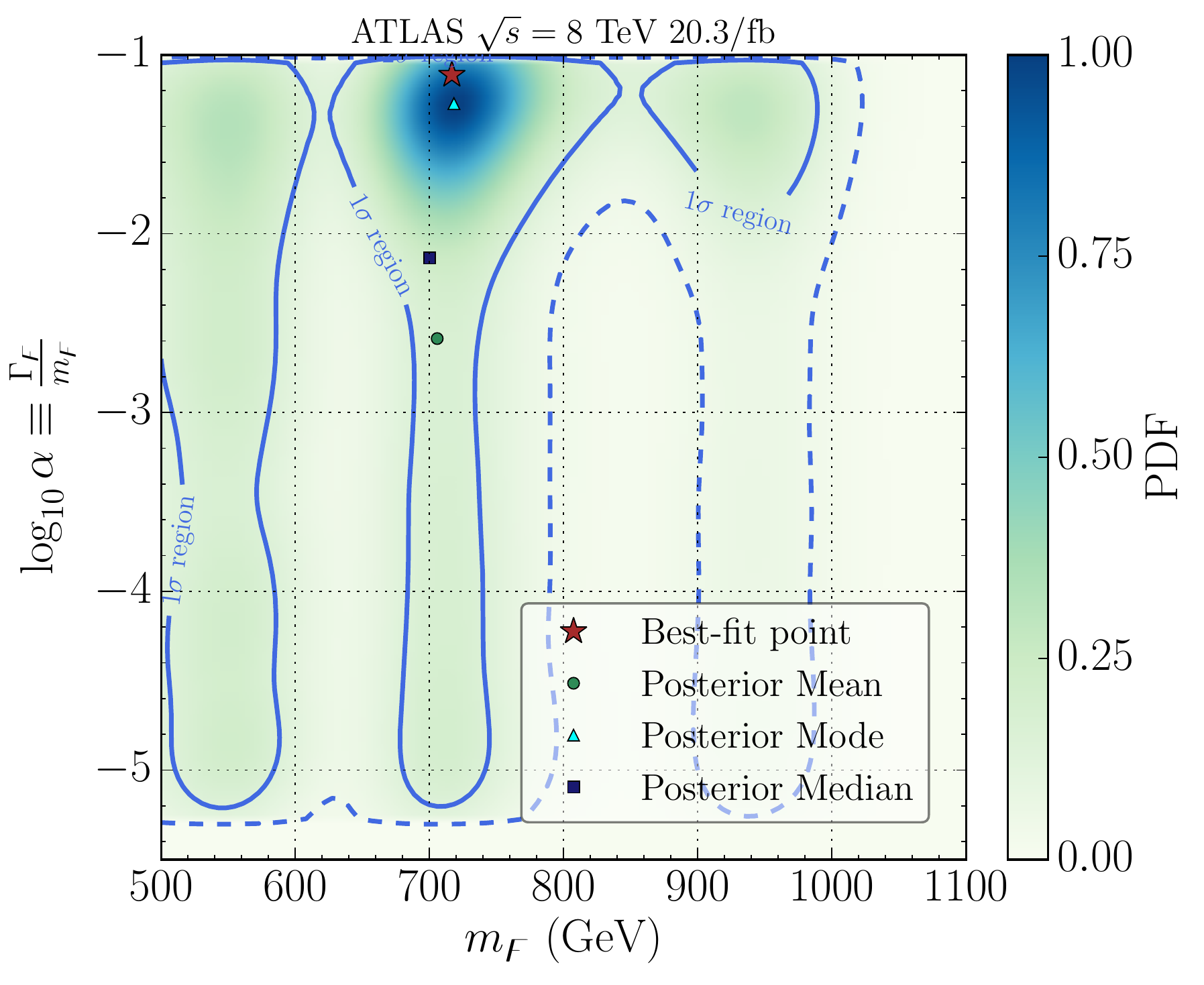}
\label{fig:8}
}
\phantom{\includegraphics[width=0.325\linewidth]{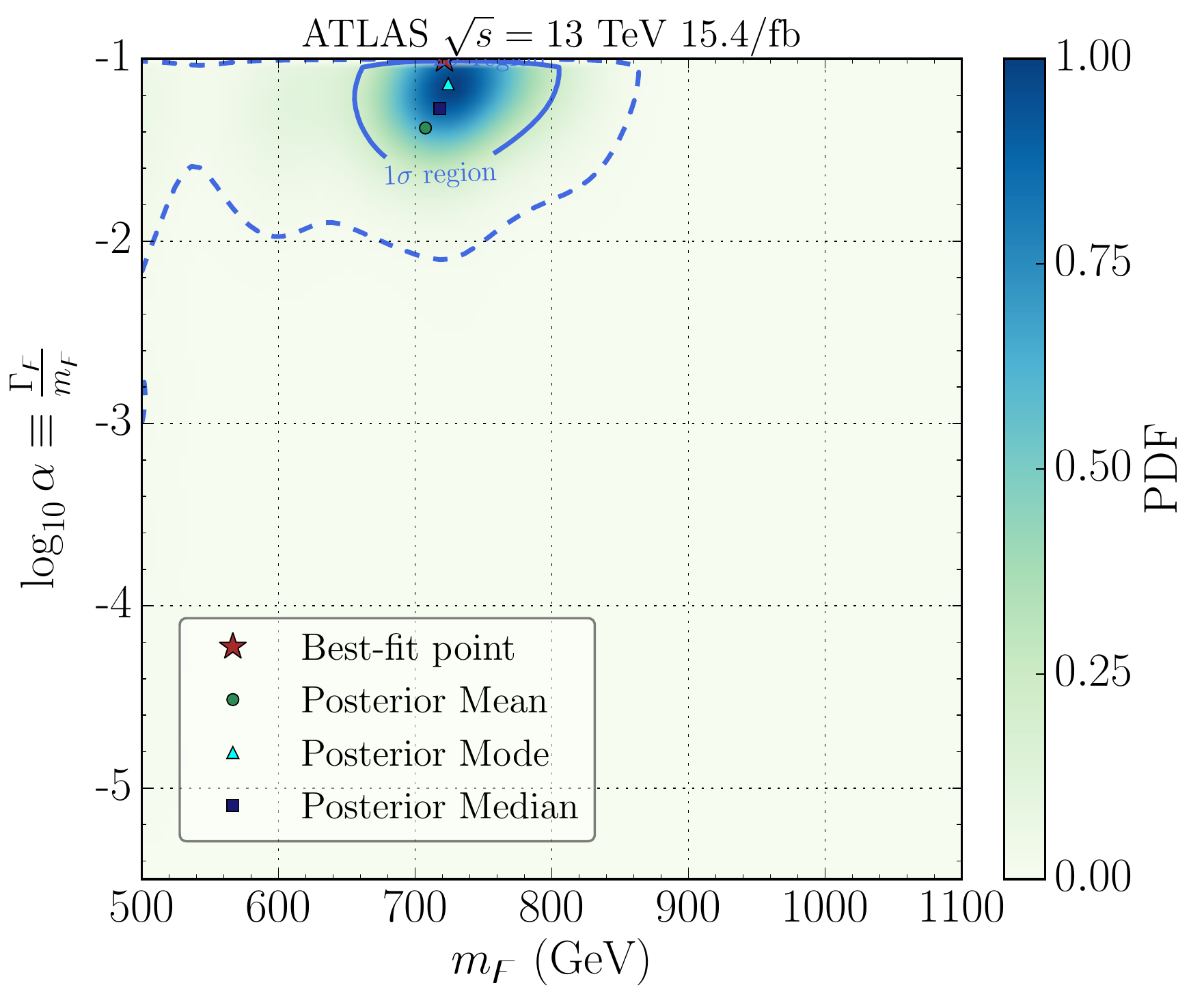}}


\subfloat[ATLAS \roots{13} $3.2\invfb$ data.]{
\setcounter{subfigure}{2}
\includegraphics[width=0.325\linewidth]{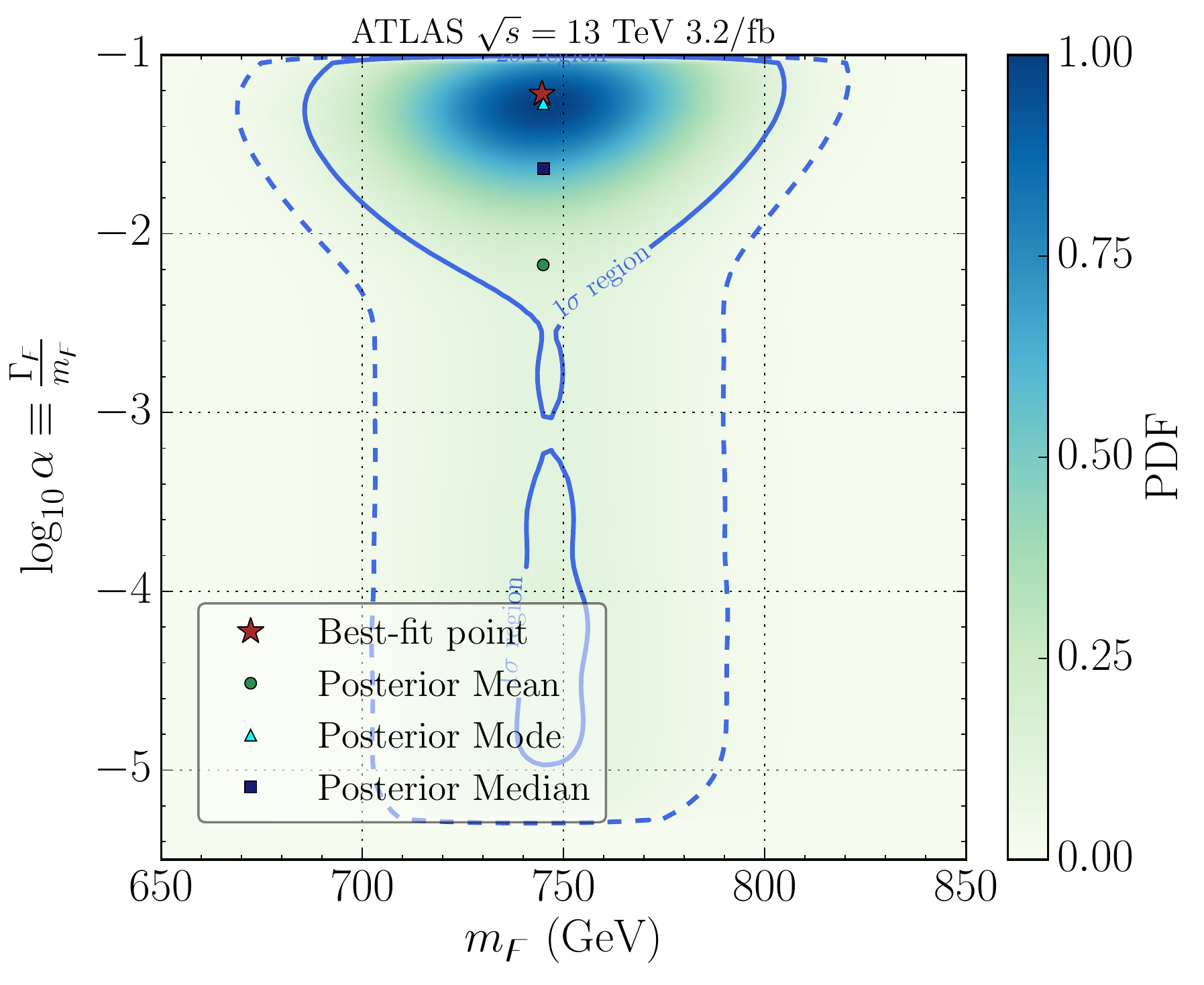}
\label{fig:13}
}
\subfloat[ATLAS \roots{13} $15.4\invfb$ data.]{
\setcounter{subfigure}{4}
\includegraphics[width=0.325\linewidth]{mass_width_pdf_13_ichep}
\label{fig:13_ichep}
}


\subfloat[ATLAS \roots{8} $20.3\invfb$ and \s{13} $3.2\invfb$ data.]{
\setcounter{subfigure}{3}
\includegraphics[width=0.325\linewidth]{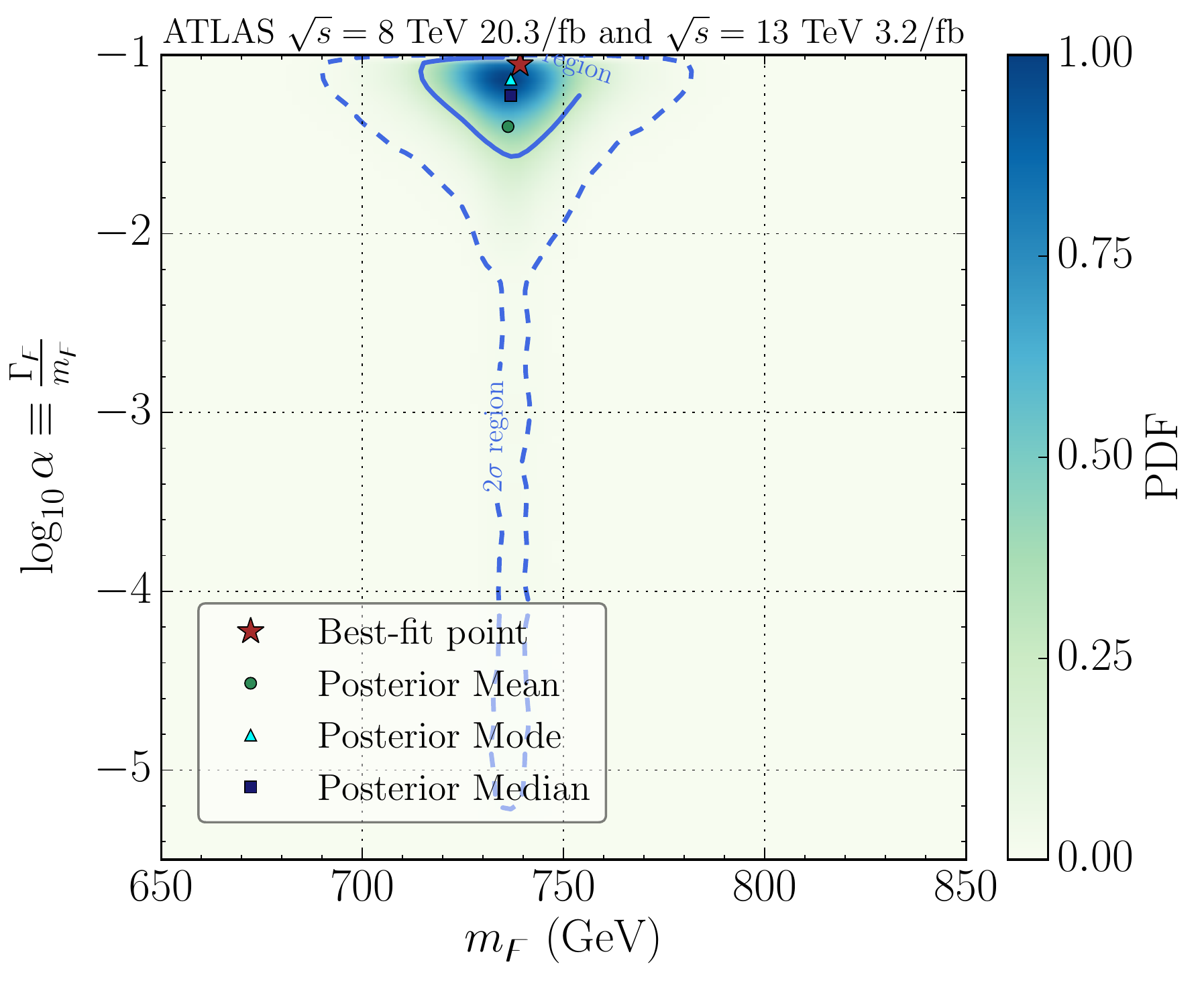}
\label{fig:combined_pdf}
}
\subfloat[ATLAS \roots{8} $20.3\invfb$ and \s{13} $15.4\invfb$ data.]{
\setcounter{subfigure}{5}
\includegraphics[width=0.325\linewidth]{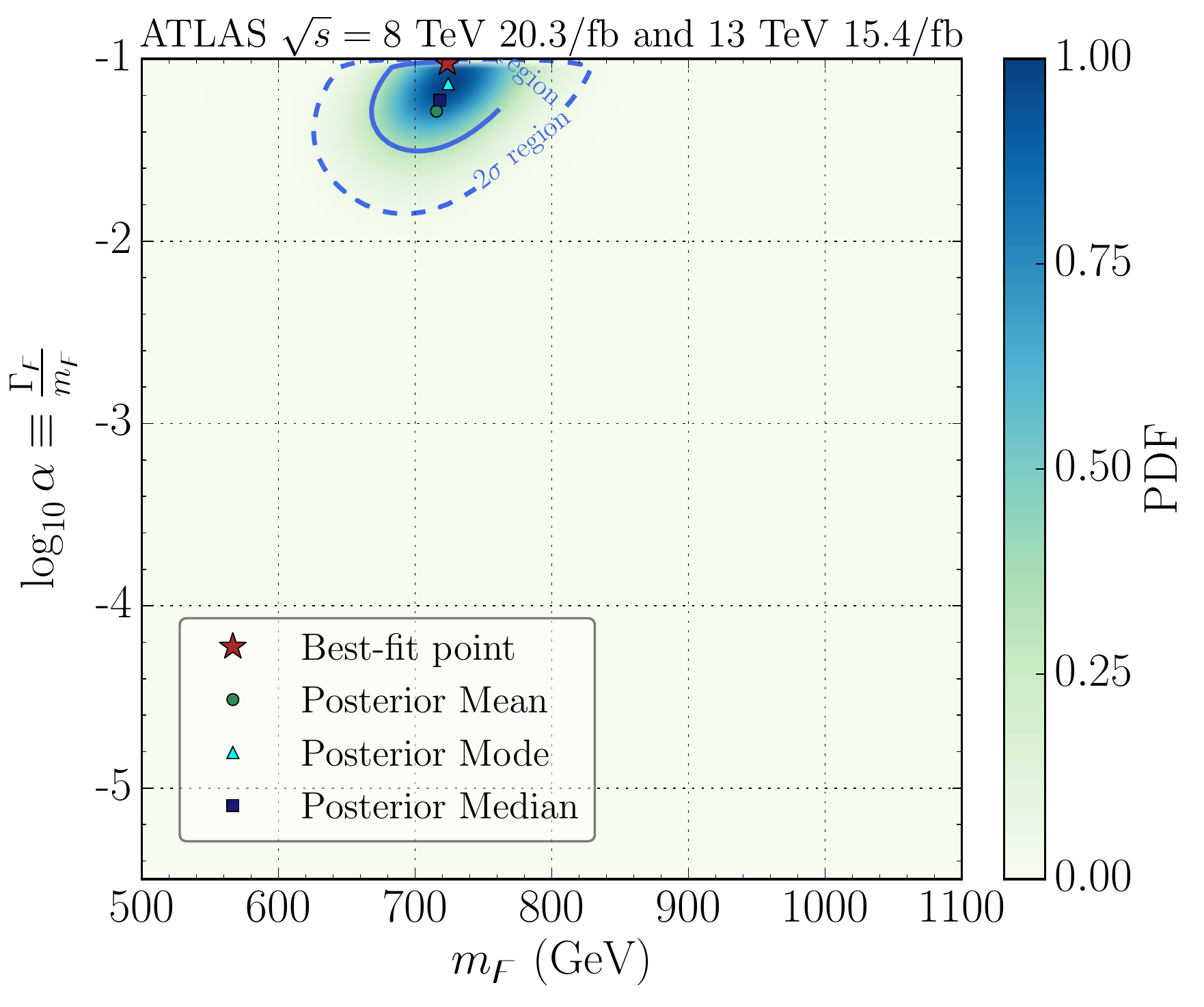}
\label{fig:combined_pdf_ichep}
}


\caption{\label{fig:pdf} The posterior pdf (see \refcite{Fowlie:2016hew}) for 
the mass and width of \digamma. Left-panels: \protect\subref{fig:8} ATLAS \roots{8} data, 
\protect\subref{fig:13} \s{13} $3.2\invfb$ data and \protect\subref{fig:combined_pdf} combined. Right panels:
\protect\subref{fig:13_ichep} \s{13} $15.4\invfb$ data and \protect\subref{fig:combined_pdf_ichep} \s{8} and \s{13} $15.4\invfb$ combined.}
\end{figure*}

We supplied the likelihood and priors to \texttt{MultiNest}\footnote{We picked an evidence tolerance of $0.01$ and $1000$ live points per dimension.}\cite{Feroz:2007kg,Feroz:2008xx,2013arXiv1306.2144F}, which performs numerical integration via the nested sampling algorithm\cite{skilling2006,Skilling:2004}, returning a Bayesian evidence, \eg
\begin{equation}\label{eq:example}
\cond{\data}{\SM} = \int \like(\bm x) \cdot \cond{\bm x}{\SM} \, d\bm x,
\end{equation}
where the factors in the integrand are the likelihood and prior, respectively, as discussed above, and $\bm x$ denotes the SM ansatz parameter set (see \refapp{app:evidences}) for a complete expression). Finally, we calculated Bayes-factors, which are ratios of Bayesian evidences. For a clear picture of the changing relative plausibility of the \digamma versus the SM, we calculated Bayes-factors for \s{8}, \s{13} $3.2\invfb$ and \s{13} $15.4\invfb$ separately and combined.  We validated our likelihood function and scanning by checking that we approximately reproduced the best-fit \digamma properties and significances reported by ATLAS; see \reffig{fig:best_fits} and \reffig{fig:pdf}. We found local significances of $3.9\sigma$ and $2.1\sigma$ at \s{13} $3.2\invfb$ and \s{8}, respectively, and a combined local significance of $4.1\sigma$.\footnote{We assumed that the log-likelihood ratio was $\frac12\chi_1^2$-distributed \see{Cowan:2010js}.} We found best-fits for $(\md, \alpha)$ at about $(745\gev, 0.06)$, $(717\gev, 0.08)$ and $(739\gev, 0.09)$ at \s{13} $3.2\invfb$, \s{8} and combined, respectively. We, furthermore, found $1.4\sigma$ tension between the preferred ratio of cross sections in \s{13} $3.2\invfb$ and \s{8} data with mass and width fixed to their \s{13} $3.2\invfb$ best-fits.
We found no indication that our signal model in \refeq{eq:signal_dist} was an inadequate or poor approximation to the unknown DSCB function.

%
%

We calculated that the \roots{8} data slightly disfavours the \digamma by a Bayes-factor of about $0.7$. At \roots{13} with $3.2\invfb$, the story changes. The \digamma is favoured by about five with the region around $\md \approx 750\gev$ dominating, as expected. We calculated that the \s{8} and \s{13} $3.2\invfb$ ATLAS data combined favoured the \digamma by about $7.7$. This is greater than a naive multiplication of Bayes-factors; the Bayes-factors cannot be combined by multiplication, as the evidences are dependent. All told, however, a combination of \s{8} and \s{13} with $15.4\invfb$ disfavours the \digamma by about $0.7$. Whilst this is evidence against the \digamma, it is ``not worth more than a bare mention'' on the Jeffreys' scale. The evidences and Bayes-factors are summarised in \reftable{tab:evidences}.


%
%

\begin{table}
\begin{ruledtabular}
\begin{tabular}{lccc} 
& \multicolumn{2}{c}{Evidence} \\
\cline{2-3}
Data     & SM & SM + \digamma & Bayes-factor\\
\hline
ATLAS $8\tev$ $20.3\invfb$ & $2.4\cdot10^{-64}$ & $1.7\cdot10^{-64}$ & $0.71$\\
ATLAS $13\tev$ $3.2\invfb$ & $6.8\cdot10^{-64}$ & $3.1\cdot10^{-63}$ & $4.6$\\
ATLAS $13\tev$ $15.4\invfb$ & $2.8\cdot10^{-87}$ & $7.2\cdot10^{-88}$ & $0.26$\\
\hline
$8\tev$ + $13\tev$ $3.2\invfb$ & $1.7\cdot10^{-127}$ & $1.3\cdot10^{-126}$ & $7.7$\\
$8\tev$ + $13\tev$ $15.4\invfb$ & $6.8\cdot10^{-151}$ & $5.0\cdot10^{-151}$ & $0.73$\\
\end{tabular}
\end{ruledtabular}
\caption{\label{tab:evidences}Evidences for the SM ansatz and \digamma resonance. Bayes-factors of greater than one indicate that the \digamma is
favoured.}
\end{table}

%
%

%
%

\subsection{Prior sensitivity of the Bayes-factor}\label{sec:priors}  

There is, of course, that thorny issue that a Bayes-factor is a functional of our priors for the models' parameters. This is most dangerous in cases in which an evidence is sensitive to the interval of a prior for a parameter. Consider a model with a single parameter $\lambda$ with a linear prior in the interval $\Delta \lambda$, and suppose that the likelihood is reasonable only in the interval $\sigma_\lambda$. The evidence would be approximately proportional to $\sigma_\lambda / \Delta \lambda$. For robust inference, it appears that we require prior information indicating a plausible interval. In fact, this is not strictly necessary. Such factors could cancel in a Bayes-factor, even for improper priors, \ie $\Delta \lambda\to\infty$. Furthermore, pragmatically, one could calculate an evidence for a sub-model with $\lambda$ in a region of interest (\eg a region searched by an experiment), $\delta\lambda$, learning whether such a sub-model was confirmed by data. This, in effect, shifts any $1/\Delta\lambda$ factors from an evidence into a prior for a sub-model, \ie 
\begin{equation}
\prob{\text{sub-model}} = \prob{\text{model}} \cdot \cond{\lambda\in\delta\lambda}{\text{model}} \propto \delta \lambda / \Delta\lambda.
\end{equation}
The resulting Bayes-factor indicates the relative change in plausibility of the sub-model. 

There may, furthermore, be cases in which our prior information cannot uniquely determine a distribution for a parameter upon an interval, \ie several choices of distribution may appear consistent with our prior information. This is problematic; the different choices may lead to different Bayes-factors. However, remarkably, in many cases particular ignorance about a parameter uniquely determines a distribution \see{jaynes2003probability}. If we are ignorant, \eg of the scale of $\lambda$, our prior should be invariant under $\lambda \to A\lambda$, leading to a logarithmic prior, $p(\log\lambda) = \text{const}$.\footnote{Whilst this distribution is improper, in practice we may be in possession of prior information that $\lambda$ cannot be arbitrarily great or small, though we are ignorant of its scale within that interval, leading to a proper distribution.}

Let us consider these issues in detail for all our priors. To quantify sensitivity, we recalculated Bayes-factors for the \s{8} + \s{13} $3.2\invfb$ data:
\begin{itemize}
\item Fortunately, because the SM background ansatz was common to each model, any such factors originating from SM background parameters would vanish in a ratio of evidences, \ie a Bayes-factor. Similarly, we anticipated limited sensitivity to the shape of the priors for the SM background ansatz parameters.

\item Whilst this was not the case for the \digamma mass and width, narrower intervals than those in \reftable{tab:priors} would imply that we were in possession of prior information that precluded resonances in regions that were searched by ATLAS. Wider intervals would merely damage the plausibility of the \digamma model by diluting its evidence, though extreme $\alpha\equiv \md / \gd$ could be implausible from the perspective of QFT. Our intervals matched the resonance masses and widths searched for by ATLAS, \ie we considered the change in plausibility of a \digamma resonance searched for by ATLAS. We picked logarithmic priors for the mass and width of the \digamma; anything else would imply prior information favouring particular scales in the intervals in \reftable{tab:priors}. 

Nevertheless, with linear priors (which imply that prior information favoured the highest scales permissible) for the \digamma mass and $\alpha \equiv \md / \gd$, we find a Bayes-factor of $39.7$ in favour of the \digamma.\footnote{This Bayes-factors was found by nested sampling. All further Bayes-factors were found by re-weighting an existing chain.} This is about $5$ times greater than previously; dominantly because a linear prior favoured the best-fitting $\alpha \approx 0.09$ by about $10$ relative to a logarithmic prior. From a theoretical perspective, however, a substantial width was, if anything, implausible relative to a narrow width; if we had any reliable prior information about the scale of the \digamma width, it was that should be narrow.

\item We picked a linear prior for the ratio of the \s{13} and \s{8} cross-sections. The interval spanned $2.5$, corresponding to a light-quark initial state, to $5$, corresponding to gluon fusion. This prior was motivated by knowledge about plausible production mechanisms; we were not ignorant of its scale and a light-quark initial state was a priori as plausible as gluon fusion. Nevertheless, we found that a logarithmic prior, which implies that prior information favoured a light-quark initial state, reduced the Bayes-factor by a factor of about $0.9$.

\item We picked a logarithmic prior between $5$ and $200$ for the number of \digamma signal events in the \s{13} $3.2\invfb$ search. Whilst we were ignorant of scale within an interval, an extreme number of signal events was implausible from the perspective of QFT (an extreme cross section) and experiments (no other evidence for a resonance with an extreme cross section), resulting in an upper limit. Reducing the maximum number of events to $50$ increased the Bayes-factor by a factor of about $1.6$. Reducing the minimum number of events would, asymptotically, result in an SM background-like model and thus a Bayes-factor of $1$.
 
\end{itemize}  
We stress that our interpretation is that our Bayes-factor is the change in relative plausibility of a \digamma resonance searched for by ATLAS, \ie with a mass in the interval $200\gev$ to $2\tev$ and $\alpha\equiv \md / \gd$ in the interval $5\cdot 10^{-6}$ to $0.1$ (see \reftable{tab:priors}), and that the priors chosen reflected knowledge about cross sections and widths in QFT, and, in some cases, ignorance of scale or location. The Bayes-factor increased by a factor of about $5$ in the worst-case  --- linear priors of the \digamma mass and width; however, it was a somewhat dishonest choice, since it implied that prior information favoured an appreciable \digamma mass and width. In other words, the prior information was sufficient to insure a weak dependence on choices of prior.

\subsection{Posterior distributions of \digamma properties}

The posterior pdfs for the \digamma properties were by-products of our calculations. Considering the combined \s{8} and \s{13} $3.2\invfb$ data, for the mass, the posterior mean, median and mode were about $\md \approx 737\gev$, with a symmetric two-tailed $68\%$ credible region spanning $724\gev$ to $747\gev$. For the width, the posterior mean, median and mode in $\log \alpha$ differed, but spanned about $\alpha \approx 0.05$ to $0.08$, with a one-tailed $1\sigma$ ($2\sigma$) lower bound at $0.05$ ($0.004$). Finally, for the ratio of cross sections, the posterior pdf favoured $\sigma_{13/8}\approx 5$, corresponding to gluon fusion, but smaller ratios were permitted with a one-tailed $1\sigma$ ($2\sigma$) lower bound at $3.8$ ($2.8$). In other words, the posterior pdf favoured a mass of about $740\gev$, a large width and production by gluon fusion, as expected. 

We show posterior pdf on the $(\md, \gd)$ plane in \reffig{fig:pdf} for \s{8}, \s{13} $3.2\invfb$ and \s{13} $15.4\invfb$ data separately and combined. The credible regions at $\md\lesssim500\gev$ (not shown) were vulnerable to digitisation errors in the data scraped from the low-mass region in \refcite{Aaboud:2016tru,ATLAS:2016eeo}.
Fortunately, that region ultimately contributed little to the evidences or our conclusions. The three prongs in the pdf at \s{8} in \reffig{fig:8} originated from deficits and excesses that surrounded the excess near $750\gev$ in \reffig{fig:digamma_8}. The pdf at \s{13} $3.2\invfb$ in \reffig{fig:13} exhibited a single prong around $750\gev$ that narrowed once data was combined in \protect\reffig{fig:combined_pdf}. 

\section{Conclusions}

Statistical anomalies near $750\gev$ in searches for diphoton resonances at the LHC resulted in a frenzy of model building. In the 
official analyses, the data was investigated with frequentist techniques. To sidestep issues regarding the interpretation of significances, with Bayesian statistics, we calculated the change in plausibility of the \digamma resonance versus the SM in light of the ATLAS  \s{8} $20.3\invfb$, \s{13} $3.2\invfb$ and \s{13} $15.4\invfb$ diphoton searches. There was limited freedom in the choice of priors for the \digamma: we matched, where possible, the ranges of width and mass searched by ATLAS. Since the models were composite, we expected limited sensitivity to the priors for the SM background ansatz and found weak dependence on our choice of priors for the \digamma signal. Ideally, we would have combined ATLAS and CMS data, though the numerous categories
in the latter make a combination challenging. Our \digamma was a toy-model described by a simple Breit-Wigner with mass, width and cross section potentially within reach of the ATLAS search; we would find different Bayes-factors for a theory that precisely predicted the \digamma properties (though know of no such theory). 

We calculated that the relative plausibility of the \digamma increased by about $7.7$ in light of the ATLAS data available at the height of the excitement. This should be contrasted with conclusions from frequentist analysis, \eg the probability of obtaining a test statistic as extreme as that observed in the \s{13} $3.2\invfb$ search were the SM correct was about $0.02$ ($2.1\sigma$). This Bayes-factor was unimpressive, especially considering that \eg SM precision measurements could quash that preference and that a width of $\gd \approx 0.06 \md$ was somewhat unexpected, a fact not reflected by our priors. On the other hand, a combination with CMS data could have increased the Bayes-factor, though there may have been tension in the preferred width. Considering all data, including \s{13} $15.4\invfb$, the \digamma was disfavoured by about $0.7$. As well as aiding our understanding of the $750\gev$ anomaly, we hope our calculations serve as a proof of principle for Bayesian model comparison and parameter inference in future experimental searches.

\appendix\section{Evidence calculation}\label{app:evidences}

An expression for an evidence was written schematically in \refeq{eq:example}. As an example, we now write in detail the evidence of the \SMdigamma\ model in light of the ATLAS \roots{13} $3.2\invfb$ data. All other evidence integrals were performed in a similar manner. We begin from the usual expression for the evidence \see{Jeffreys:1939xee},
\begin{align}
\begin{split}
\mathcal{Z} & \equiv \cond{\text{ATLAS \roots{13} $3.2\invfb$}}{\SMdigamma} \\
&= \int \like(\bm y) \cdot \cond{\bm y}{\SMdigamma} \, d\bm y,
\end{split}
\end{align}
where $\bm y$ denotes $\{a_{13}, b_{13}, n_{b13}, n_{s13}, \md, \alpha\}$, \ie the \SMdigamma\ parameters, the likelihood function, $\like(\bm y)$, was a product of Poissons (\refeq{eq:like}) and the priors were independent. Explicitly,
\begin{equation}
\cdots  = \int \prod_i \frac{\lambda_i(\bm y)^{n_i} e^{-\lambda_i(\bm y)}}{n_i!} \cdot \prod_j \cond{y_j}{\SMdigamma} \, d y_j,
\end{equation}
where $i$ indexes diphoton mass bins in the ATLAS \roots{13} $3.2\invfb$ search, such that $n_i$ is the number of observed events in bin-$i$, and $j$ indexes the \SMdigamma\ parameters. The expected number of events per bin, $\lambda_i$, was a function of the \SMdigamma\ parameters,
\begin{equation}\label{eq:bin_int}
\lambda_i(\bm y) = \int\limits_{\text{bin-$i$}} n_{b13} \cdot p_b(m_{\gamma\gamma}; \bm y) + n_{s13} \cdot p_s(m_{\gamma\gamma}; \bm y) \, d m_{\gamma\gamma},
\end{equation}
where the $p_b$ and $p_s$ are the background and signal diphoton distributions in \refeq{eq:bg_dist} and \refeq{eq:signal_dist}, respectively. The product of priors was
\begin{equation}
\prod_j \cond{y_j}{\SMdigamma} \propto \frac{1}{n_{s13} n_{b13}} \frac{1}{\md \alpha}
\end{equation}
inside the intervals in \reftable{tab:priors} and zero elsewhere. The reciprocal factors were for logarithmic priors. The priors were, of course, normalised such that
\begin{equation}
\int \prod_j \cond{y_j}{\SMdigamma} \,dy_j = 1.
\end{equation}
The integration in \refeq{eq:bin_int} was performed analytically; all other integration was performed numerically with the nested sampling Monte-Carlo algorithm\cite{skilling2006,Skilling:2004}.

\bibliographystyle{h-physrev.bst}
\bibliography{digamma}
\end{document}